\documentclass[pra,twocolumn,showpacs,preprintnumbers,floatfix]{revtex4-1}
\usepackage{graphicx}% Include figure files
\usepackage{dcolumn}% Align table columns on decimal point
\usepackage{bm}% bold math
\usepackage{latexsym,epsfig}
\usepackage{graphicx}
\usepackage{verbatim}
\usepackage{comment}
\usepackage{amsmath}
\usepackage{amssymb}
\usepackage{stmaryrd}
\usepackage{color}

\newcommand{\beq}{\begin{equation}}
\newcommand{\eeq}{\end{equation}}
\newcommand{\bea}{\begin{eqnarray}}
\newcommand{\eea}{\end{eqnarray}}
\newcommand{\ben}{\begin{eqnarray*}}
\newcommand{\een}{\end{eqnarray*}}
\newcommand{\bfig}{\begin{figure}}
\newcommand{\efig}{\end{figure}}

\begin{document}

\title{Supersolid in a one dimensional model of hardcore bosons}

\author{Tapan Mishra$^1$, Ramesh V. Pai$^2$ and Subroto Mukerjee$^{3,4}$}
\affiliation{$^1$International Center for Theoretical Sciences, TIFR, Bangalore, 560 012, India}
\affiliation{$^2$ Department of Physics, Goa University, Taleigao Plateau, Goa 403 206, India}
\affiliation{$^3$ Department of Physics, Indian Institute of Science,
Bangalore, 560 012, India}
\affiliation{$^4$ Centre for Quantum Information and Quantum Computing (CQIQC), Indian Institute of Science,
Bangalore, 560 012, India}

\date{\today}

\begin{abstract}
We study a system of hardcore boson on a one-dimensional lattice with frustrated next-nearest
neighbor hopping and nearest neighbor interaction.
At half filling, for equal magnitude of nearest and next-nearest neighbor hopping, the ground state of this
system exhibits a first order phase transition from a Bond-Ordered (BO) solid to a Charge-Density-Wave(CDW) solid as a
function of the nearest neighbor interaction. Moving away from half filling we 
investigate the system at incommensurate densities, where we find a SuperSolid (SS) phase which has concurrent off-diagonal 
long range order and density wave order which is unusual in a system of hardcore bosons in one dimension.
Using the finite-size Density-Matrix Renormalization Group (DMRG) method, 
we obtain the complete phase diagram for this model.

\end{abstract}

%\pacs{03.75.Lm, 05.10.Cc, 05.30.Jp} \keywords{Suggested keywords}

\maketitle

%%%%%%%%%%%%%%%%%%%%%%%%%%%%%%%%%%%%%%%%%%%%%%%%%%%%%%%%%%%%%%%%%
\section{INTRODUCTION}
\label{sect:intro}

Supersolid phases of matter which feature both off diagonal superfluid order and 
long range crystalline order have been a subject of
intense research in the last decade. While there is still no clear evidence for 
the occurrence of this phase in solid Helium~\cite{kim_2012}, there are proposals 
for creating such a state in optical lattices of cold atoms~\cite{kellmann_2009}. 
Model Hamiltonians which describe supersolid phases exist, of which one of the most 
rigourously studied is one of hardcore bosons on a lattice with further 
neighbor interactions~\cite{herbert_2001,batrouni_ss,boninsegni,wessel,damle,melko}.
Supersolid phases have also been predicted in models of softcore bosons and 
binary mixtures\cite{batrouni_prl_2005,mishrass,mishra_mixture,batrouni_bose_fermi}
and quantum spin systems\cite{lee,mila}. It has been proposed that a system of polar gases in 
optical lattices is a suitable test bed to observe this exotic phase of matter. 
Pioneering experiments on chromium Bose-Einstein condensates~(BECs)~\cite{Griesmaier2005} have been recently followed by the realization of
quantum gases in other highly-magnetic species, including dysprosium Bose and Fermi gases~\cite{Lu2011} and erbium condensates~\cite{Aikawa2012}.
Significantly more dipolar gases may be realized by means of polar molecules, which have large electric dipole moments of the order of the Debye or larger.
Seminal experiments on KRb molecules at JILA~\cite{Ni2008} has opened the door towards achieving a quantum degenerate gas of polar molecules, 
and various experimental groups worldwide are currently involved in this enterprise~\cite{Wu2012,Takekoshi2012}. Rydberg gases constitute yet another 
possible realization of highly dipolar gases~\cite{Gallagher2008}. The successful manipulation of polar lattice gases in optical lattice experiments could lead to the
observation of supersolid phases.

On the other hand, the ability to produce frustration in optical lattices of cold atoms 
has opened up possibilities to realize interesting superfluid and Mott states 
which have additional kinds of order arising from the kinetic 
frustration~\cite{dhar1,dhar2,santos_cmi}. Kinetic frustration in these systems 
is produced by the competition of two different hopping processes from a site to 
different sites with different signs of the hopping amplitude. 
The two different
hopping processes could be to a nearest neighbor site and
a next-nearest neighbor site. It is thus interesting to study the interplay between nearest neighbour interaction and kinetic 
frustration away from commensurate densities, which can potentialy stabilize supersolid phase.
\bfig[!t]
  \centering
  \includegraphics*[width=0.45\textwidth,draft=false]{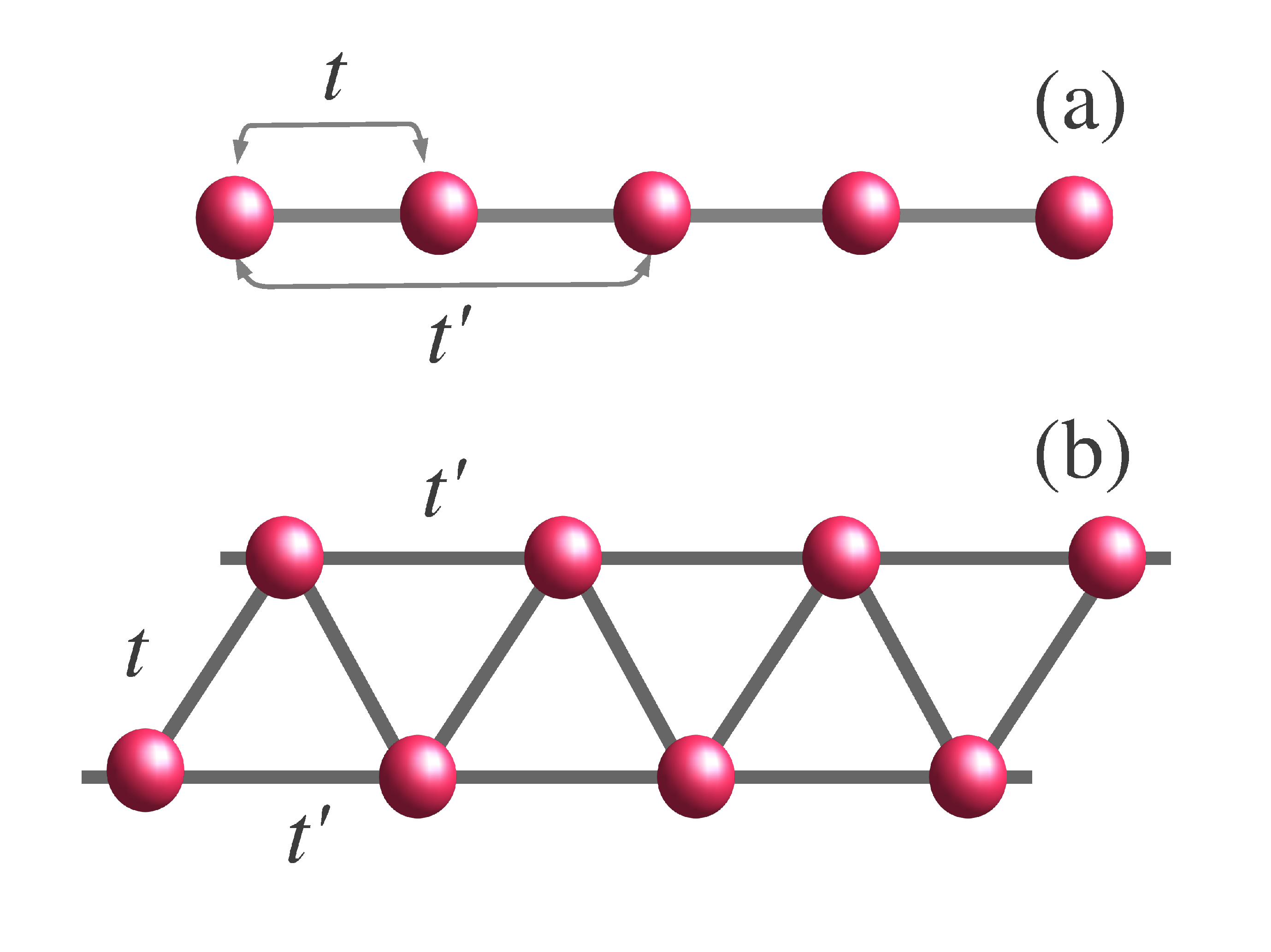}
    \caption{(Color online) (a)$1D$ lattice with nearest and next-nearest neighbour hopping. (b) Equivalent triangular ladder geometry. 
    The arrows are the representation of the hopping directions.}
    \label{fig:ladder}
\efig

In this paper, we study such a model of hard core bosons hopping on a one-dimensional 
lattice with nearest neighbor hopping and interaction and next-nearest neighbor 
hopping that induces kinetic frustration as shown in Fig.~\ref{fig:ladder}(a). This model is equivalent to a system of triangular ladder as shown in 
Fig.~\ref{fig:ladder}(b).
The model describing such a system 
can be described by the Hamiltonian

\begin{eqnarray}
\!\!\!\!\!\! H \! &=& \!  - t \sum_i (a_{i}^{\dagger}a_{i+1}^{\phantom \dagger} + \text{h.c.})
 \! - \! t' \sum_i (a_{i}^{\dagger}a_{i+2}^{\phantom \dagger}+\text{h.c.}) \notag \\
 &+& \sum_i V \left(n_i-\frac{1}{2}\right) \left(n_{i+1}-\frac{1}{2}\right)
\label{eq:ham}
\end{eqnarray}
where $a_i^{\dagger}$ and $a_i^{\phantom \dagger}$ are creation and annihilation operators
for hard core bosons
at site $i$, and $n_i=a_i^{\dagger}a_i^{\phantom \dagger}$ is the boson number operator
at site $i$. $t$ and $t'$ are the nearest
and next-nearest neighbor hopping amplitudes. $V$ represents the nearest neighbor repulsion. 
Frustration in this model is introduced by choosing $t > 0$ and $t' < 0$.
The model described by Eq.~\ref{eq:ham} can also be thought of as a triangular 
ladder where the nearest neighbor hopping and interaction 
are along the rungs and the next-nearest neighbor hopping is
along the legs. In this work we scale the energies with respect to $t$ by considering $t=1$, therefore, all the parameters considered are 
dimensionless.  
As discussed in Ref. \cite{mishra_ttpv}, this model does not have a simple representation in terms of spinless fermions due to the
presence of the next-nearest-neighbour hopping term. At half filling and for $V=0$, apart from the trivial point $t'=0$, there exists one more point 
corresponding to $t'=-t/2$, where the exact ground state can be obtained~\cite{mishra_ttpv}.

The model of Eq.~\ref{eq:ham} has been studied recently by us at 
half filling~\cite{mishra_ttpv}. The ground state phase diagram has three different phases,
a uniform superfluid (SF), an insulating charge
density wave (CDW) crystal and a bond ordered insulator (BO). 
When $t=|t'|$, only the insulating (gapped) phases occur and there is a first order transition between them as
a function of $V$. In this paper we study the system by doping 
it away from half filling to see what types of gapless phases might arise. 
By performing a detail analysis using the FS-DMRG method, we obtain
a complete ground state phase diagram for this model. 
Our main result is that in addition to the gapped CDW and BO phases, 
the phase diagram contains a gapless supersolid(SS) phase, which is a phase with concurrent 
superfluid and charge density wave order. This is summarized in the 
phase diagram of  Fig.~\ref{fig:phasedia}, which is plotted as a function of the 
chemical potential $\mu$ and interaction $V$.
In the following sections, we present details of our calculations
\bfig[!t]
  \centering
  \includegraphics*[width=0.45\textwidth,draft=false]{fig2.eps}
    \caption{(Color online)Phase diagram of the $t-t'-V$ model at 
commensurate and incommensurate with $t'=-t$ in the chemical 
potential and interaction $(\mu,V)$ plane. 
Both $\mu$ and $V$ are in units of $t$ and $\mu=0$ corresponds 
to the case of half filling where only the CDW and BO phases occur. 
For other values of $\mu$, we obtain the additional supersolid phase SS. 
The phase boundaries marked with green circles are those between 
gapped and gapless phases and are obtained by calculating the charge gap. 
The phase boundaries marked with red triangles and blue squares with error bars for some values of $V$ are between the gapless superfluid and the 
supersolid phases and are determined by looking at the diverging compressibility and the structure factor respectively as explained in the text.}
    \label{fig:phasedia}
\efig
\section{Details of the DMRG method}
We study the model described by Eq.~\ref{eq:ham} using the finite-size DMRG method with open boundary
conditions~\cite{white_92,schollwock_review_05}. This method is best
suited for (quasi-)one-dimensional problems~\cite{schollwock_review_05}. 
For most of our calculations we study system sizes up to 200 sites and retain up to $256$ density
matrix eigenstates with the weight of the discarded states in the density matrix less than $10^{-6}$.
We compute various physical quantities to characterize the different phases. 
Some of these quantities have been calculated by us using the DMRG method to study 
related models~\cite{mishra_ttpv,tapan_tvvp}. 
We describe below the quantities which are most important for the 
characterization of the different phases.

In order to distinguish between gapped and gapless phases we calculate the chemical potentials
\beq
\mu=(\mu^++\mu^-)/2
\label{eq:gap}
\eeq
where $\mu^+=E(N+1,L)-E(N,L)$ and $\mu^- = E(N,L)-E(N-1,L)$.
In Eq.~\eqref{eq:gap},
$E(L,N)$ is the ground-state energy of the system with $L$ sites and $N$ bosons.

The CDW order in the system can be quantified by calculating the structure factor,
which is the Fourier transform of the density-density correlation function
\begin{equation}
 S(k)=\frac{1}{L^2}\sum_{i,j}{e^{ik(i-j)}\langle{n_{i}n_{j}}\rangle }%-\langle n_i\rangle\langle n_j
%\rangle)}.
\label{eq:str}
\end{equation}

The BO phase is characterized by a non-zero value of the bond-order parameter
\beq
O_{BO}=\frac{1}{L}\sum_i(-1)^i B_i,
\label{eq:obow}
\eeq
where
\beq
B_i=\langle a_i^\dagger a_{i+1}^{\phantom \dagger}+a_{i+1}^\dagger a_i^{\phantom \dagger}\rangle.
\eeq

\section{Results and discussion}
We first discuss how to obtain the signature of the gapless and 
gapped phases at incommensurate and commensurate densities respectively. 
This is done by computing the chemical potential $\mu$ defined in Eq.~\ref{eq:gap} 
for various densities $\rho$.
We start at a value of $\rho$ far away from half filling
and then dope the system to increase $\rho$ gradually. Since the model 
considered is particle-hole symmetric, the signatures at
densities above half filling are mirror reflections of
those below half filling. 
The gapless to gapped transition can be seen in the $\rho - \mu$ plot as shown in Fig.~(\ref{fig:rhomu}).
It can be seen that there exists a jump in $\mu$ as a function of $\rho$
at $\rho=0.5$ for different values of $V$. 
The corresponding length of the plateau in $\rho$ decreases as 
$V\sim 3.0$ where the gap exists only very close to 
the transition between CDW and BO at $\mu=0$ and again increases. The end points of the plateaux
trace out the BO nd CDW phases which are shown in Fig.~(\ref{fig:phasedia}). The BO and CDW phases are characterised by the finite bond order parameter and 
density wave structure factor as defined in Eq.~\ref{eq:obow} and Eq.~\ref{eq:str} respectively. 
\bfig[!b]
  \centering
  \includegraphics*[width=0.45\textwidth,draft=false]{fig3.eps}
    \caption{(Color online)(a),(b),(c),(d) show the $\rho$ vs. $\mu$ plotted for 
different values of $V$. The appearance of a plateau indicates an incompressible 
(gapped) phase. The length of the plateau measured along the $\mu$ direction 
gives the size of the gapped region in the phase diagram at a given $V$. 
At $V=3.0$, which is where the direct transition from CDW to BO occurs 
at $\mu=0$, the gap exists only very close to the transition point and hence there is no plateau.}
    \label{fig:rhomu}
\efig

It is obvious from the Fig.~(\ref{fig:rhomu}) that the compressibility
$\partial \rho /\partial \mu$ is zero along the plateau and is finite on the shoulders around the plateau. 
However, it should be noted that there is a
kink in the $\rho$ vs. $\mu$ plot for $V \geq 2.0$ ,
where the chemical potential tends to saturate with respect to $\rho$ and therefore the compressibility diverges.
These kinks appear for all the values of $V\geq 2.0$ considered in our calculation. The divergent 
compressibility can be regarded as the signature of a phase transition which can be located from 
the kink position. This phase transition corresponds to the transition among the gapless 
phases, the SS and the SF. Once, $\rho$ is increased beyond the position 
of the kink, $\mu$ increases monotonically with $\rho$ indicating a finite 
compressibility in the gapless SS phase. The kink positions which give us 
the phase boundaries between the gapless phases are shown in Fig.~(\ref{fig:phasedia}). 
Note that we cannot characterize the nature of these gapless phases (i.e. say whether 
they are SS or SF) from the above analysis. For that we require to calculate 
the correct order parameters as well, which we discuss in the following sections.
\bfig[!t]
  \centering
  \includegraphics*[width=0.45\textwidth,draft=false]{fig4.eps}
    \caption{(Color online)On-site number density $\langle n_i \rangle$ 
plotted as a function of site index $i$ for different values of $\rho$ with (a)$\rho=0.50$,(b)$\rho=0.49$,
    (c)$\rho=0.48$, (d)$\rho=0.47$ and (e)$\rho=0.46$ for $V=4.0$. It can be seen that while there is 
perfect CDW order at $\rho=0.5$, there is a modulation over at a wave vector $k_m$ that appears
as one moves away from $\rho=0.5$.}
    \label{fig:ni}
\efig
\bfig[!b]
  \centering
  \includegraphics*[width=0.45\textwidth,draft=false]{fig5.eps}
    \caption{(Color online)$S(k)$ vs $k$ for different densities for 
$V=4.0$. It can be seen that the location of the peak shifts away 
from $k=\pi$ as the filling $\rho$ decreases for $\rho=0.5$. The shift is found to be linear in $\rho$ such that the modulation vector $k_m=2\pi \rho$.}
    \label{fig:sk}
\efig
The CDW and BO phases which occur at $\mu=0$ are gapped and thus one would expect them to remain 
robust to small changes in $\mu$. Thus, we would expect the CDW and BO phases to appear as lobes (as seen in Fig.~(\ref{fig:phasedia})). 
To understand what happens as we move away from half filling, we calculate
the density-density structure factor as defined in Eq.~\ref{eq:str} and also 
look at the local density $n_i$ as a function of lattice site $i$. It can 
be seen from Fig.~(\ref{fig:ni}) that there is a modulation of the density 
with wavevector $k_m$ superimposed on the CDW order as the filling is 
changed from $\rho=0.5$. We can quantify the dependence of $k_m$ on $\rho$ by plotting the 
density-density structure factor $S(k)$ as in Fig.~(\ref{fig:sk}). With a modulation $k_m$ in the density, 
$S(k)$ has a peak at $k=\pi-k_m$, which can be seen as the peak shifts away from $k=\pi$ as $\rho$ changes. 
Tracking the positions of the peaks yields $k_m=2\pi \rho$. Similar feature has been studied before in a system of hardcore 
bosons in zig-zag ladder~\cite{rossini}. From Fig.~(\ref{fig:rhomu}), it can be seen that 
the state, one obtains for $\rho \neq 0.5$, where the density modulation occurs is a compressible (gapless) 
state and thus corresponds to a supersolid (SS). However, it is a supersolid, where the charge ordering wavevector 
is dependent on the filling $\rho$.

\bfig[!t]
  \centering
  \includegraphics*[width=0.45\textwidth,draft=false]{fig6.eps}
    \caption{(Color online)$S(k=2\pi\rho)$ is plotted as a function of 
$V$ for different $\rho$. The peak position drifts linearly with the filling fraction $\rho$. The modulation wavevector $k_m=2\pi \rho$.}
    \label{fig:str}
\efig
\bfig[!b]
  \centering
  \includegraphics*[width=0.45\textwidth,draft=false]{fig7.eps}
    \caption{(Color online)$S(\pi-k_m)$ vs $V$ for $\rho=0.46$ for different system sizes shows that 
there is a fairly steep jump at a particular value of $V$ which does not appear to drift appreciably 
with increasing system size. However, the jump seems to be come more gradual with increasing system size.}
    \label{fig:str46}
\efig

The SS phase shares phase boundaries with the CDW, BO phase 
and SF phases as can be seen from Fig.~(\ref{fig:phasedia}). The phase boundary between the SS and gapped CDW and BO phases is obtained 
by measuring the charge gap which is zero in the SS phase but finite in the gapped phases. The other phase boundary between the SS 
and the SF phases cannot be obtained in this way since they are all gapless. To obtain this boundary we plot 
the peak value 
of the structure factor $S(\pi-k_m)$ as a function of $V$ for different values of $\rho$ as shown in Fig.~(\ref{fig:str}). 
The value of $S(\pi-k_m)$ at each density as a function of $V$ shows a fairly steep increase at a particular value of $V$. This 
value of $V$ at which this happens does not drift appreciably with increasing system as shown in Fig.~(\ref{fig:str46}) although 
it appears steeper for smaller system sizes. The phase boundary obtained this way coincides with the one obtained from the 
positions of the kinks in the plots of Fig.~(\ref{fig:rhomu}) as discussed earlier. This validates this particular way of 
obtaining the phase boundary. The lines of constant density in the SS and SF phases of the phase diagram can be seen 
in Fig.~(\ref{fig:zoom}).

\bfig[!t]
  \centering
  \includegraphics*[width=0.45\textwidth,draft=false]{fig8.eps}
    \caption{(Color online) An enlarged view of the SS lobe and regions around it of the phase 
diagram of Fig.~(\ref{fig:phasedia}), with the lines of constant density marked.}
    \label{fig:zoom}
\efig

The SF phase might be stabilized over a larger part 
of the phase diagram and be easier to detect if we choose a different set of parameters, 
say $|t'| < t$.  As we have seen in our previous work, for such a choice of parameters, 
it is possible to obtain a regular superfluid phase even exactly at half-filling~\cite{mishra_ttpv} 
and it is quite likely that this phase will remain over a fairly large part of the phase diagram even when 
we move away from half-filling. However, for these parameters, 
it is likely that the SS phase will occupy a smaller region of the phase diagram.

\section{Conclusions}
\label{sect:conc}
We have studied a system of hardcore bosons in a one dimensional 
optical lattice with frustrated next-nearest-neighbour hopping and nearest neighbour interaction. 
Using the finite-size DMRG method we have obtained the ground state phase 
diagram of this model and shown that in addition to gapped CDW and BO phases, 
it also displays the regular SS phase, which has concurrent 
superfluid and CDW order.
\section{Acknowledgments}
We would like to thank Arun Paramekanti and B. P. Das, Thierry Giamarchi and 
Abhishek Dhar for many useful discussions. 
SM thanks the Department of Science and Technology, Govt. of India for support. 
RVP thanks the UGC, Govt. of India for support.

\end{document}